# Evidence of Ballistic Thermal Transport in Lithium Niobate at Room Temperature

R. A. Pulavarthy and M. A. Haque[a]

*Department of Mechanical & Nuclear Engineering, the Pennsylvania State University, University Park, PA 16802, USA*

In ballistic transport, heat carriers such as phonons travel through the solid without any scattering or interaction. Therefore, there is no temperature gradient in the solid, which seems to transport the heat without getting heated itself. Ballistic transport is typically seen in high purity crystals at either temperatures below ~10 K, or physical size below ~100 nm, where the mean free path of the carrier is larger than the solid itself. In this letter, we show evidence of ballistic transport at room temperature in lithium niobate wafers in the in-plane and cross-plane directions under both steady state and high frequency heating that are monitored using both infrared and resistance thermometry. We report phonon mean free path in lithium niobate around 425 microns, which is about 50 times higher than the largest phonon mean free path in the literature at room temperature. Above this length-scale, temperature gradient gradually emerges and the material shows completely diffusive, bulk transport at about 4 mm length. Our observations will impact phonon-based electronics such as thermal transistor, thermal logic gate and memory currently impossible at room temperature. If 1 micron electron mean free path in graphene gives the highest-mobility (>200,000 cm$^2$ V$^{-1}$ s$^{-1}$), the 425 μm mean free path of phonons in this research may realize phononics without any need for nanoscale size or ultra-cold temperatures.

---

[a] Author to whom correspondence should be addressed. Electronic mail: mah37@psu.edu, tel: 814-865-4248, fax: 814 865-9693

In dielectric and semiconductors solids, heat is mainly conducted by phonons, which interact with other phonons, defects and interfaces. The average distance they travel between each scattering event is called the mean free path, which is typically on the order of ~100 nm. Phonon transport in these solids has three possible mechanisms: a) by ballistic transport, where the phonons travel through the solid without any interaction, b) by second sound, where the energy transport is a wave like phenomenon and the phonon momentum is conserved during most of the scattering process and c) by diffusion, which is observed in almost all the bulk materials. The ballistic transport and second sound are observed at very low temperatures and in materials of high purity. When the characteristic size of the solid is much larger than the mean free path of the phonon, there is massive scattering and the heat flow is diffusive. In nano-systems where the solid size is lesser or comparable to the phonon mean free path, heat conduction contains a major ballistic part. Studies have shown ballistic-diffusive phonon transport at room temperature in nanostructures [1-3] and two-dimensional materials like graphene [4]. To date, the largest phonon mean free path at room temperature is reported to be around 8.3 microns in SiGe nanowires[5].

Fourier's law of heat conduction, given by Equation 1, assumes local thermal equilibrium and the heat flow to be diffusive in nature. In a solid with thermal conductivity κ, it defines the heat flux vector **q** is proportional to the temperature gradient ∇T in the material. Derived from the kinetic theory, the thermal conductivity is related to other physical properties as in Equation 2.

$$\mathbf{q} = -\kappa \, \nabla T \qquad (1)$$

$$\kappa = \frac{1}{3} C v \lambda \qquad (2)$$

where, C is the specific heat per unit volume of the material, v is the speed of sound in the material and λ is the phonon mean free path. In the case of ballistic phonon transport, the Fourier law breaks down and is known to over-predict the thermal transport [6]. In such cases, the Boltzmann Transport Equation (BTE), given below by Equation 3, has to be solved to study the thermal transport.

$$\frac{\partial f}{\partial t} = \mathbf{v_g} \cdot \nabla f + \frac{\partial \mathbf{k}}{\partial t} \cdot \nabla_k f = \left|\frac{\partial f}{\partial t}\right|_{collision} \qquad (3)$$

where, f is the phonon distribution function dependent on position vector **r**, time t, wave vector **k** of the phonons. **v_g** is the group velocity vector of phonons which is defined as

$$\mathbf{v_g} = \frac{\partial \omega}{\partial \mathbf{k}} \tag{4}$$

where, ω is the phonon frequency. The right hand side of Equation 3 denotes the change in the phonon distribution due to the collisions. There are several approaches to solve the BTE using various approximations since a direct analytical solution to its raw form is not practical. Few of them include the relaxation time approximation [7], neglecting the optical phonons, isotopic approximation [8] and gray approximation where the relaxation time and phonon velocity are taken to be frequency independent. Literature exists on computational studies of thermal transport in ballistic regime[2,9]. It is widely accepted that phonons have majority states with mean free paths in the order of ~100 nm. Despite that, a minority of phonon states exist with mean free path as big as 10 μm at room temperature and still have a strong contribution to thermal conductivity of the solid[10].

There is a very little body of experimental evidence of long mean free paths at the meso or macro scales at room temperature. Nanostructures and cryogenic temperatures have a limited utility to applications. In this letter, we present the experimental observation of ballistic phonon transport at room temperature in a bulk lithium niobate wafer. Lithium niobate is a dielectric material with attractive piezoelectric, optical, electro-optic and photorefractive properties. It has non-linear optical properties which makes it an encouraging material for conversion of thermal radiation[11]. It is also cheaply available with a very high curie temperature (~1500 K) [12] that makes it favorable for applications in high temperature devices. To demonstrate ballistic thermal transport at room temperature, we performed steady state thermal transport experiments along in-plane and cross-plane directions, followed by high frequency heat pulse experiments typically used to capture non-diffusive transport. We then estimated the phonon mean free path by measuring thermal conductivity as function of specimen length. We also performed finite element simulation to highlight the drastic difference between our experimental results and the conventional Fourier transport.

In-plane ballistic transport: To map the in-plane thermal transport, we patterned 30 microns wide nickel lines in serpentine shape to act as a thin film heater. This is shown in Figure 1a. Next to this heater, several thin nickel film lines were patterned in parallel to act as heat-sinks so that resistance thermometry can be used to measure the local temperature for cross-validation of the infrared thermometry. Figure 1a shows the infrared (IR) image after passing current through the thin film heater. This excites the phonons in that region that propagate until they scatter at a boundary such as the heat-sink films. To study the experimentally obtained temperature domain, we developed a multi-physics model incorporating Joule heating and diffusive heat transfer using the same geometry and heating current input to the commercially available COMSOL software. The simulation results are shown in Figure 2b. The stark contrast between the two temperature profiles is highlighted in Figure 2c, which shows the temperature line scan along x-x′ direction. It shows that diffusive heat transfer would be localized in the heater area, gradually decreasing to the room temperature. Given the low thermal conductivity (~ 4 W/m-K) of the lithium niobate substrate, very little heat is expected to be transported to the heat sinks, indicated by the clear absence of any temperature peaks on the metal heat-sink lines. However, the IR images capture a completely different physics, where the heater lines show high temperature spikes but the substrate between two heater lines do not heat up appreciably. To highlight that substrate carries the heat without itself getting hot, we locate 4 locations on the substrate marked as A, B, C and D in Figure 2c. Intriguingly, the substrate temperature is close to room temperature right next to the heater (point A) and it remains fairly constant through the points B, C and D. However, very distinct temperature peaks are seen in the metal heat sink lines next to these locations. These observations clearly show that the phonons propagate in a ballistic manner, i.e., they do not create temperature gradient unless they scatter at the boundary (which is the interface between metal and the substrate).

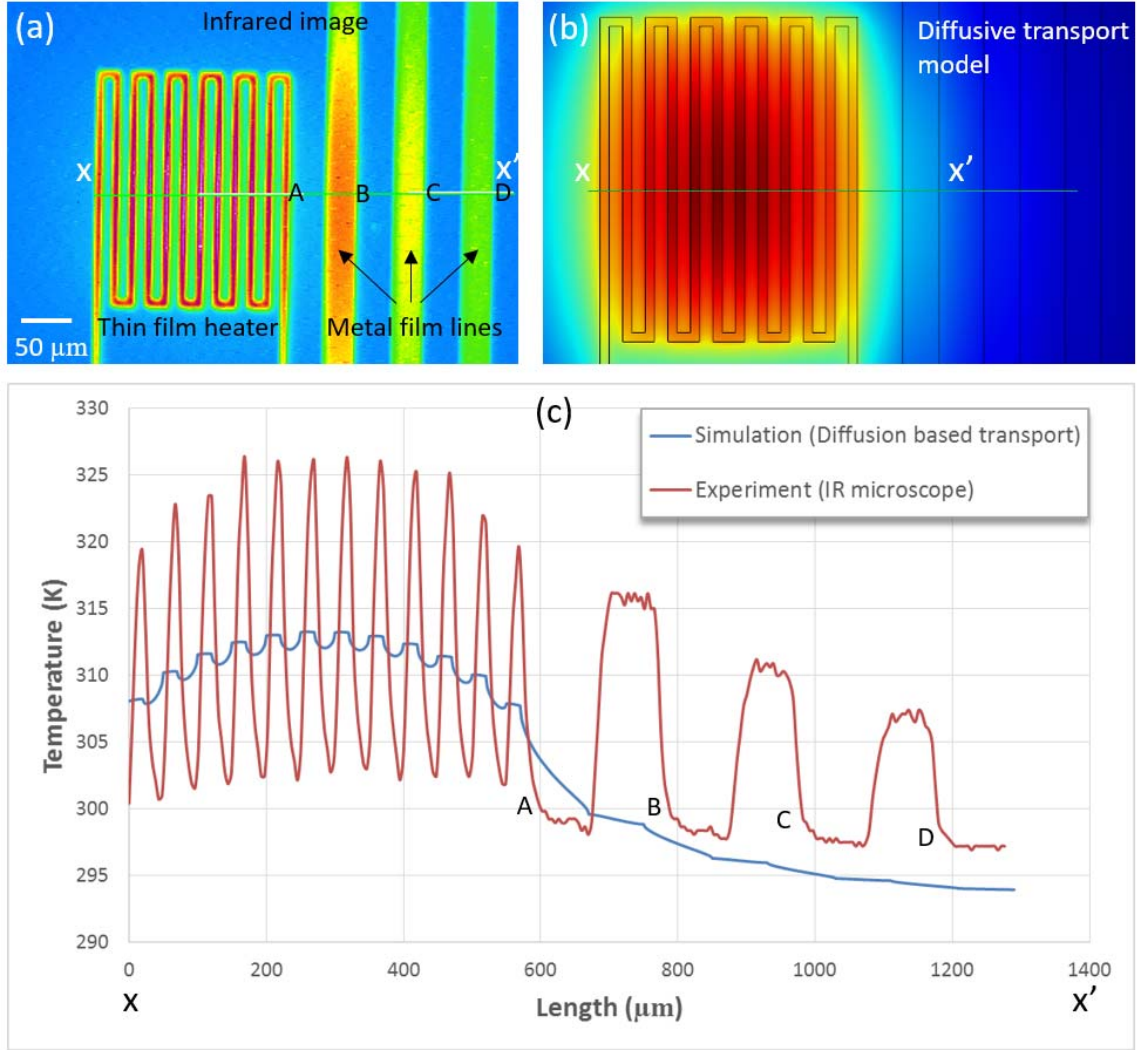

FIG. 1. Experimental results on in-plane thermal transport compared with diffusive, finite element simulation, (a) infrared image of the lithium niobate substrate after activating the thin film heater (b) Diffusive mode finite element simulation using exactly same geometry and heating current, (c) temperature profiles along the scan lines x-x′.

Cross-plane ballistic transport: To demonstrate cross-plane ballistic transport, we mount a 2mm x 2mm ceramic heater on one side of the lithium niobate wafer and observe the temperature profile along the thickness direction. Figure 2a shows a cross-sectional image from IR microscope, which shows a sharp temperature drop at the interface after which the temperature profile is uniform inside the lithium niobate. This is better seen in Figure 2b that shows the line scan results on the temperature. Once again, the lithium niobate wafer carried the heat without getting hot, however a small temperature spike (marked by the oval) was seen at the boundary of the solid. This is due to energy dissipation when the phonons scatter at the substrate boundary, which reconfirms the ballistic transport. Figure 2c shows the simulation results from COMSOL for the same operating and boundary conditions. Here, the diffusive heat transfer develops a small temperature gradient along the thickness direction. This is because the thin glass substrate heats up due to phonon scattering, which is absent in the experimental observations (Figure 2b). To revalidate the signature of ballistic transport, we modified this cross-plane experiment by attaching a second material (a Kapton foil) to the lithium niobate. As expected, the niobate substrate carried the heat to the Kapton foil to raise its temperature to nearly the magnitude of the heater itself, while it remained at a lower temperature.

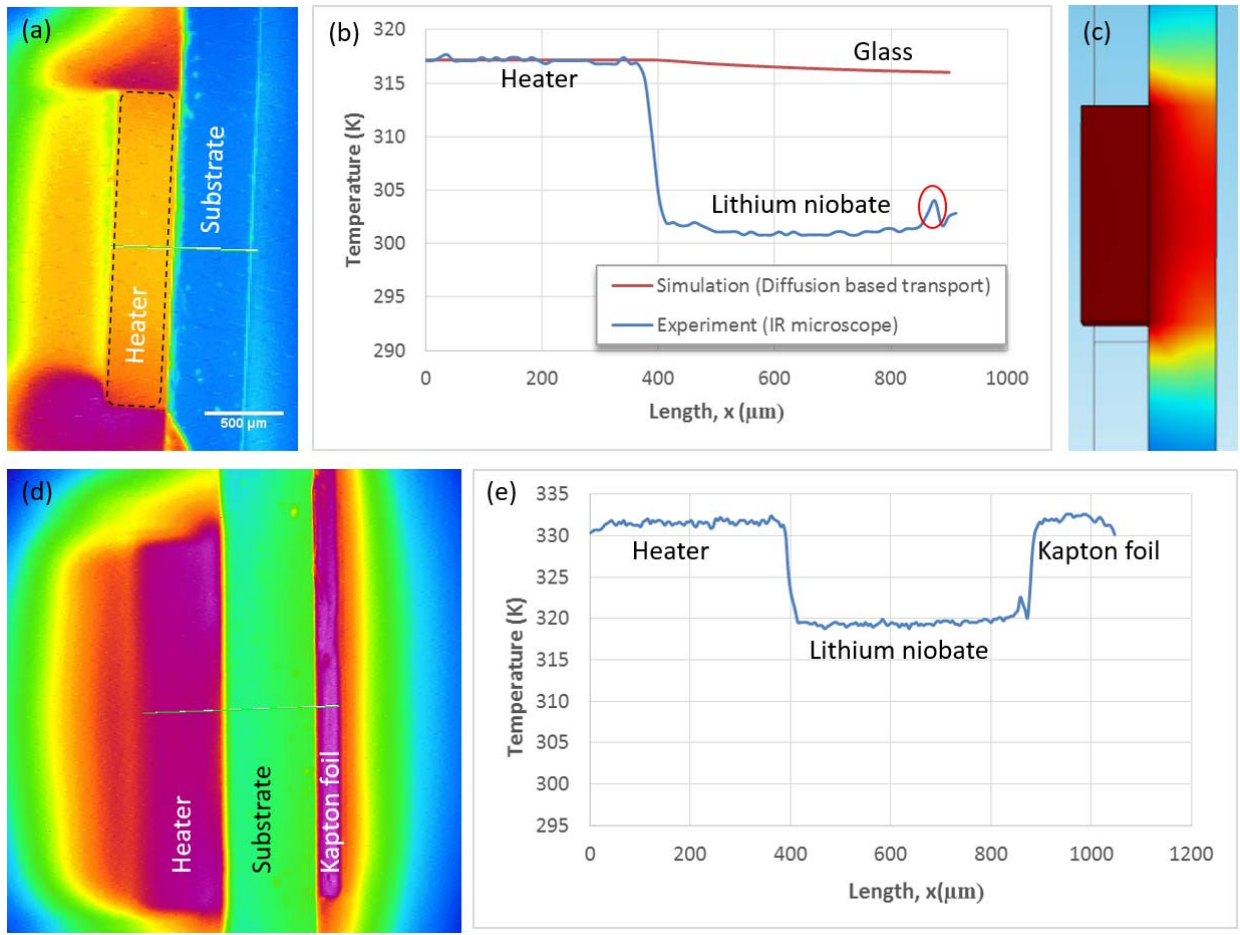

FIG. 2. Experimental results on cross-plane thermal transport, (a) IR image showing ceramic heater on the lithium niobate substrate, (b) temperature profile along the line scan in (a), (c) comparison of temperature profile obtained from experiment and diffusive heat transfer simulation (d) IR image confirming ballistic transport in the substrate to heat up a second material (Kapton foil) through boundary scattering (e) temperature profile of the IR image in (d).

Transient Heat Pulse Response: Classical studies[13] on ballistic transport have used transient heat pulse on one side of the solid while capturing the thermal response on the other side as function of time using a thin film bolometer. Figure 3a shows the landmark results on highly pure NaF crystals at cryogenic temperatures,[14] where two distinct response peaks appear. The first peak (marked as L) has smaller amplitude and originates from longitudinal phonons while the significantly higher second peak (marked as T) is originated from transverse phonons. At room temperature, the peaks flatten dramatically to a single waveform. To capture this transient phenomenon, we performed heat pulse experiments on the setup shown in Figure 3b. We passed 8 milli-amp current pulses with 2 millisecond duration through the metallic film heater. The thermal pulses travel through the material and is sensed by the metal line similar to resistance thermometry. Figures 3c and 3d show the input and output pulse profiles respectively. Interestingly, we observed a sharp single peak resembling the longitudinal phonon mode with very distinct suppression of the transverse mode. This is qualitatively depicted in Figure 3d, where the thermal signatures appear not only ballistic but also single phonon mode. Both longitudinal and transverse modes can be dissipative, but the longitudinal mode is relatively less dissipative than the transverse mode [15].

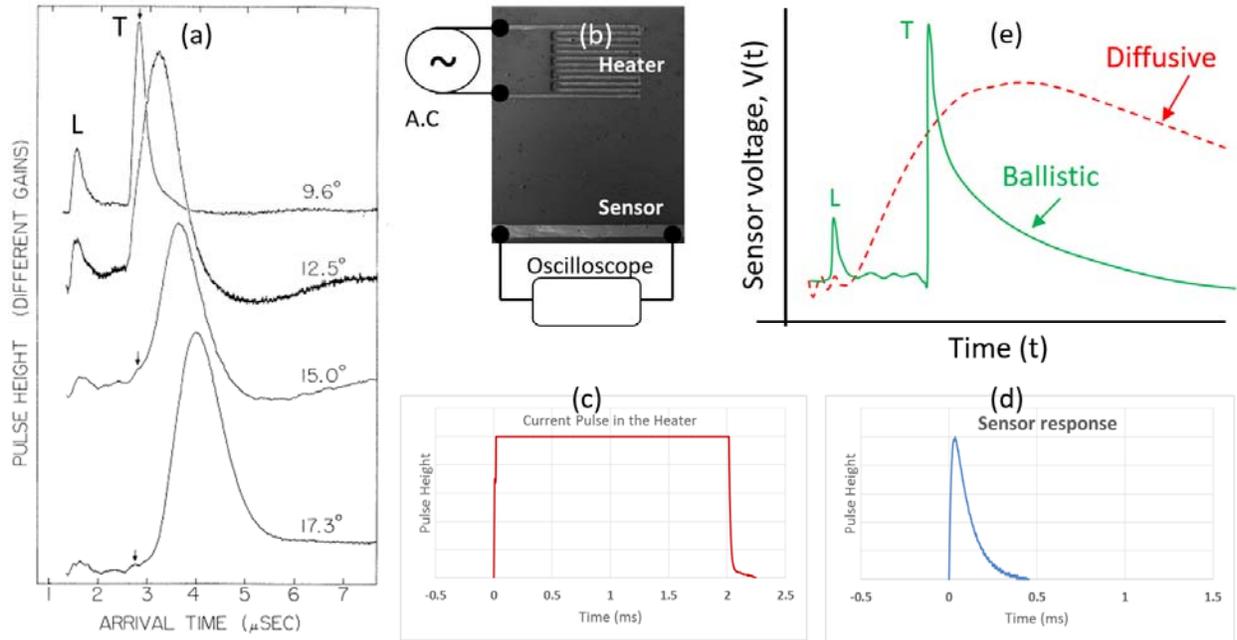

FIG. 3. (a) Signature of ballistic transport in heat pulse experiments on highly pure NaF[14], (b) experimental setup for this study (c) input current pulse profile (d) corresponding output response profile showing single sharp peak (e) output profile qualitatively compared with ballistic and diffusing signature profiles. Item (a) is reprinted with permission from H. E. Jackson, C. T. Walker, and T. F. McNelly, Physical Review Letter, Vol. 25, Issue. 1, pp. 26-28, Copyright (1970) by the American Physical Society.

To estimate the mean free path, we adopt the procedure that measures thermal conductivity as function of the characteristic length.[1,5] From a classical point of view, ballistic transport should offer no resistance to heat conduction. But, according to the Landauer formalism quantum confinement restricts the thermal conductance in a channel due to the geometric boundaries. In this study, the lithium niobate wafer is diced to a 500 μm X 500 μm cross-section with an aspect-ratio long enough to approximate the phonon transport to be one-dimensional. One end of the specimen is fixed with the test rig while the other end is brought into mechanical and thermal (using carbon based adhesive) contact by manipulating a large material slab. The basic concept is to controllably and precisely vary the specimen length while measuring the thermal conductivity. The experimental setup is shown in Figure 4a. The freestanding part of the sample between the two ends is considered the length, L, which is varied in this experiment. A 2 mm x 2 mm ceramic heater is brought in contact to one end of the specimen to establish a temperature gradient in the specimen, which is imaged by an infrared microscope. An energy balance approach assuming the specimen to be a one-dimensional rectangular fin results in the following expression,

$$\frac{T(x)-T_\infty}{T_b-T_\infty} = \frac{\frac{T_L-T_\infty}{T_b-T_\infty}\sinh(mx)+\sinh(m(L-x))}{\sinh(mL)} \qquad (5)$$

where, $T(x)$ is the temperature at x along the length of sample specimen, $T_b$ is the base temperature at x=0, $T_L$ is the tip temperature at x=L, $T_\infty$ is the ambient temperature and m is expressed as

$$m = \sqrt{\frac{hP}{\kappa A_c}} \qquad (6)$$

where, h is the overall heat transfer coefficient, P is the perimeter and $A_c$ the cross-sectional area of the sample. h and κ are the only two unknowns in Equations 5 and 6. The evaluation of h is discussed in detail by Alam et al.[16] Using the piezo-controlled stage, we vary the specimen length till a measurable temperature gradient is seen. The corresponding length is noted as $L_0$, which is the estimated phonon mean free path. The thermal conductivity is estimated by the parametric variation of κ to get the best fit of model prediction (Equation 5) to the experimental temperature readings using the method of least squares. This is repeated at eight different lengths of the sample viz. $L_0$ to $L_7$. Figures 4b and 4c show the IR images of the sample heated at lengths $L_0$ and $L_7$ respectively. The standard of the fit between the model prediction and the experimental data is shown in Figures 4d and 4e for the corresponding values of thermal conductivity. Figure 4f shows how we use the intersection of the curve to estimate the phonon mean free path. The phonon mean free path estimated from this study is around $\Lambda = 425$ microns. When the specimen length is about 4 mm, the thermal conductivity reaches 75% of the bulk value and we did not extend the experiments.

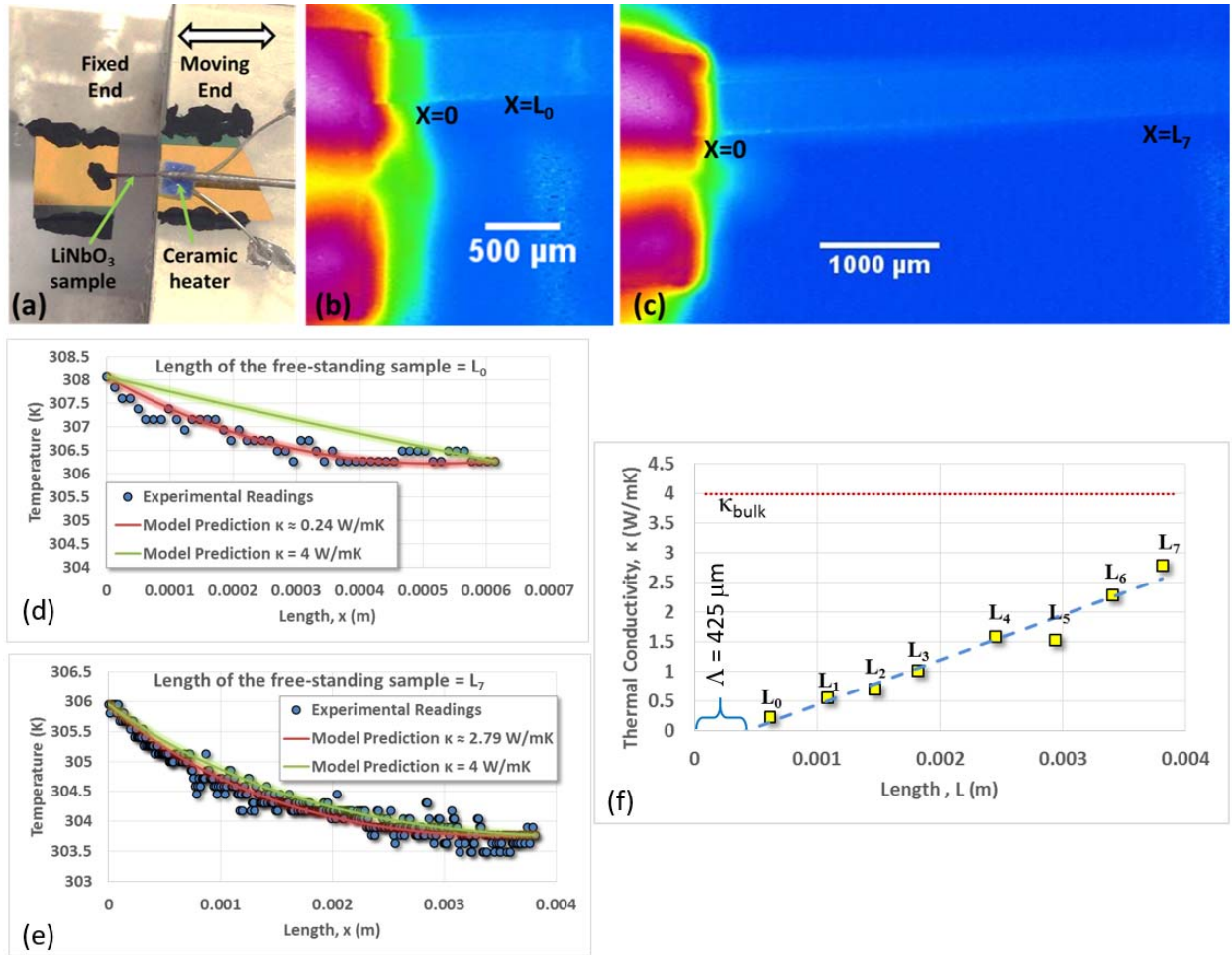

FIG. 4. Estimation of the phonon mean free path, (a) experimental set-up utilizing a piezo-controlled manipulator to control the freestanding length of a specimen (b & c) IR micrographs of the heated sample at length $L=L_0$ and $L=L_7$ respectively, (d & e) fit of the model prediction to the experimental temperature profiles at lengths $L_0$ and $L_7$, (f) estimation of phonon mean free path ($\Lambda$) from thermal conductivity variation with specimen length.

Both computational and experimental studies have explored phonon transport in the ballistic regime[17]. It is generally accepted that not all phonons contribute equally to the heat transport and low frequency phonons contribute substantially to heat transport. The phonon characteristics of LiNbO$_3$ and the influence of composition has been well established using techniques like Raman spectroscopy [18], IR spectroscopy [19]. The phonon density of states (DOS) has also been calculated using ab initio method [20]. The total density of states display the predominance in the low-frequency region since niobium and lithium atoms possess large atomic mass and small force constants respectively. It is the oxygen atoms that majorly occupy the high-frequency region. Scattering phenomenon is known to have a big role in filtering or suppressing the contribution from high-frequency phonons. Alloy scattering[5], defect scattering[21] are known to enhance this suppression. Lithium niobate is recognized to have a high density of sub-microscopic defects that are intrinsic[22] some of which may or may not depend on the crystal composition. We hypothesize this high defect concentration is responsible for the restraining of high-frequency phonons, which is supported by the strong phonon focusing in lithium niobate[23]. In the Raman spectrum, only transverse phonons are detected in one geometry while both transverse and longitudinal phonons are detected in other geometries [18]. This argument is further strengthened by the display of only one sharp peak shown in Figure 3e, in the heat pulse experiment. Efforts are currently underway to study these hypotheses and provide fundamental understanding behind the observed ballistic thermal transport. The impact of this study is highlighted by the fact that experimental research on ballistic transport has been limited by the restriction of temperature around 4 K or size below 100 nm. Discovery of room temperature ballistic transport in a macroscopic material, as demonstrated in this study, may accelerate development of phonon based electronics, which has been impossible at room temperature.

**ACKNOWLEDGMENTS**

We gratefully acknowledge support from the National Science Foundation of the USA (CMMI 1029935).

**REFERENCES**

[1]   Jaeho Lee, Jongwoo Lim, and Peidong Yang, Nano Letters **15** (5), 3273 (2015).
[2]   Y. Dong, B. Y. Cao, and Z. Y. Guo, Physica E-Low-Dimensional Systems & Nanostructures **66**, 1 (2015).
[3]   Mark E. Siemens, Qing Li, Ronggui Yang, Keith A. Nelson, Erik H. Anderson, Margaret M. Murnane, and Henry C. Kapteyn, Nat Mater **9** (1), 26 (2010).
[4]   M. E. Pumarol, M. C. Rosamond, P. Tovee, M. C. Petty, D. A. Zeze, V. Falko, and O. V. Kolosov, Nano Letters **12** (6), 2906 (2012); M. H. Bae, Z. Y. Li, Z. Aksamija, P. N. Martin, F. Xiong, Z. Y. Ong, I. Knezevic, and E. Pop, Nature Communications **4** (2013).
[5]   T. K. Hsiao, H. K. Chang, S. C. Liou, M. W. Chu, S. C. Lee, and C. W. Chang, Nature Nanotechnology **8** (7), 534 (2013).
[6]   G. Chen, Journal of Heat Transfer-Transactions of the Asme **118** (3), 539 (1996); G. D. Mahan and Francisco Claro, Physical Review B **38** (3), 1963 (1988).
[7]   D. P. Sellan, J. E. Turney, A. J. H. McGaughey, and C. H. Amon, Journal of Applied Physics **108** (11), 113524 (2010).
[8]   Javier V. Goicochea, Marcela Madrid, and Cristina Amon, Journal of Heat Transfer **132** (1), 012401 (2009).
[9]   T. Kuhn and I. J. Maasilta, in *12th International Conference on Phonon Scattering in Condensed Matter*, edited by B. Perrin, B. Bonello, A. Devos et al. (Iop Publishing Ltd, Bristol, 2007), Vol. 92; Jesse Maassen and Mark Lundstrom, Journal of Applied Physics **117** (3), 035104 (2015); G. Chen, Physical Review Letters **86** (11), 2297 (2001); Yanbao Ma, International Journal of Heat and Mass Transfer **66**, 592 (2013).
[10]  A. S. Henry and G. Chen, Journal of Computational and Theoretical Nanoscience **5** (2), 141 (2008).
[11]  A. V. Syuy, M. N. Litvinova, P. S. Goncharova, N. V. Sidorov, M. N. Palatnikov, V. V. Krishtop, and V. V. Likhtin, Technical Physics **58** (5), 730 (2013).
[12]  K. J. Kirk, N. Schmarje, and S. Cochran, Insight **46** (11), 662 (2004).
[13]  SJ Rogers, Physical Review B **3** (4), 1440 (1971); W. Dreyer and H. Struchtrup, Continuum Mechanics and Thermodynamics **5** (1), 3 (1993).
[14]  Howard E. Jackson, Charles T. Walker, and Thomas F. McNelly, Physical Review Letters **25** (1), 26 (1970).
[15]  D. Y. Tzou, Journal of Heat Transfer **136** (4), 042401 (2014).
[16]  M. T. Alam, A. P. Raghu, M. A. Hague, C. Muratore, and A. A. Voevodin, International Journal of Thermal Sciences **73**, 1 (2013).
[17]  Jeremy A. Johnson, A. A. Maznev, John Cuffe, Jeffrey K. Eliason, Austin J. Minnich, Timothy Kehoe, Clivia M. Sotomayor Torres, Gang Chen, and Keith A. Nelson, Physical Review Letters **110** (2), 025901 (2013); D. A. Broido, M. Malorny, G. Birner, N. Mingo, and D. A. Stewart, Applied Physics Letters **91** (23) (2007); L. Lindsay, D. A. Broido, and T. L. Reinecke, Physical Review Letters **111** (2), 025901 (2013); Y. S. Ju and K. E. Goodson, Applied Physics Letters **74** (20), 3005 (1999).
[18]  A. Ridah, P. Bourson, M. D. Fontana, and G. Malovichko, Journal of Physics-Condensed Matter **9** (44), 9687 (1997).
[19]  J. L. Servoin and F. Gervais, Solid State Communications **31** (5), 387 (1979).
[20]  K. Parlinski and Z. Q. Li, Physical Review B **61** (1), 272 (2000).

[21] G. H. Zhu, H. Lee, Y. C. Lan, X. W. Wang, G. Joshi, D. Z. Wang, J. Yang, D. Vashaee, H. Guilbert, A. Pillitteri, M. S. Dresselhaus, G. Chen, and Z. F. Ren,  Physical Review Letters **102** (19) (2009).
[22] H. X. Xu, D. Lee, J. He, S. B. Sinnott, V. Gopalan, V. Dierolf, and S. R. Phillpot,  Physical Review B **78** (17) (2008);   G Malovichko, V Grachev, and O Schirmer,  Applied Physics B **68** (5), 785 (1999);    SC Abrahams and P Marsh,  Acta Crystallographica Section B: Structural Science **42** (1), 61 (1986).
[23] G. L. Koos and J. P. Wolfe,  Physical Review B **30** (6), 3470 (1984).